\documentclass[conference]{IEEEtran}

\usepackage{cite}      

\usepackage{graphicx}  

\usepackage{psfrag}    

\usepackage{subfigure} 

\usepackage{url}       

\usepackage{amsmath}   
\usepackage{units}
\usepackage{booktabs}

\begin{document}

\title{ Simulation of  up- and down-going neutrino induced showers
  at the site of the Pierre Auger Observatory}

\author{\authorblockN{Dariusz G\'ora}
\authorblockA{Institut f\"ur Experimentelle Kernphysik \\
 Universit\"at  Karlsruhe, \\
  D-76021 Karlsruhe, Germany \\
   Institute of Nuclear Physics PAN \\
  ul. Radzikowskiego 152, \\
  31-342 Krak\'ow, Poland  \\
Email: Dariusz.Gora@ik.fzk.de}
\and
\authorblockN{Markus Roth}
\authorblockA{Institut f\"ur Kernphysik,\\
 Forschungszentrum Karlsruhe, \\
  D-76021 Karlsruhe, Germany \\
Email: Markus.Roth@ik.fzk.de}
\and
\authorblockN{Alessio Tamburro}
\authorblockA{Institut f\"ur Experimentelle Kernphysik \\
 Universit\"at  Karlsruhe, \\
  D-76021 Karlsruhe, Germany \\
  Email: Alessio.Tamburro@ik.fzk.de
  }}


%


\maketitle

\begin{abstract}
We present a study about the possibility to detect neutrino induced extensive
 air showers at the Pierre Auger Observatory.
 The Monte Carlo simulations performed take into account
 the details of the neutrino propagation inside the Earth, the air
 as well as  the surrounding mountains
  which are modelled by a digital elevation map.
 Details on the sensitivity with respect to the incoming direction as well
 as the aperture and the total observable event rates are calculated
 on the basis of various assumptions of the incoming neutrino flux.
\end{abstract}

%
\IEEEpeerreviewmaketitle
%
\section{Introduction}
\label{intro} 
The detection  of very high energy cosmic neutrinos, above \unit[1]{EeV}, 
is important as it may allow to identify the most  powerful sources in the
Universe. 

In the first place, neutrinos, due to their small cross-section 
can travel cosmological distances without interactions. Hence
they might carry astrophysical information about their sources which 
are commonly believed to be optically thick~\cite{Bahcall:1999yr}. 

Second, due to their connection to the emission of cosmic nuclei and gamma
rays, they might help to solve the problem of the origin of ultra high energy
cosmic rays (UHECRs). 
In fact although the existence of UHECRs is experimentally proven, their
composition and origin are still unknown. Many models have been proposed to
explain the origin of UHECRs. Some of these, which involve mechanisms of
particle acceleration ({\em bottom-up}), claim that they might be produced
by Active Galactic Nuclei (AGN) and Gamma Ray Bursts
(GRB)~\cite{Olinto:2000sa}. In such scenarios neutrinos could be produced with
an  upper flux limit given by the Waxman-Bahcall (WB)
bound~\cite{Bahcall:1999yr}. 
{ 
Other models claim that UHECRs might come from the  decay of super-massive
objects ({\em top-down}): These objects are expected to be produced by
radiation, interaction  or collapse of {\em topological defects \/} such as
monopoles, cosmic strings, etc.~\cite{Bhattacharjee:1998qc}. The topological
defect models predict a larger flux of photons and neutrinos arriving at Earth
than the {\em bottom-up \/} models. 
}
Finally, high energy neutrinos could be also produced through pion decay when 
protons interact with the cosmic microwave background. These neutrinos are
so-called {\em cosmogenic neutrinos}~\cite{cosmo}.

Nevertheless, due to their vacuum oscillations~\cite{Fukuda::2001yr}, a flux
of high energy cosmic 
neutrinos is expected to be almost equally distributed among the three
neutrino flavours. Therefore  the study of oscillation effects on high energy
neutrino fluxes can be used to study the neutrino mixing and distinguish
amongst different mass schemes.  

According to the models described above, a low incoming flux of neutrinos is
expected. In addition, neutrinos have a small interaction probability.
Therefore, in order to get a detectable rate, very large  neutrino detectors
are needed such as the Pierre Auger Observatory~\cite{Abraham:2004dt}.
In particular, for Auger Observatory   Earth-skimming $\nu_{\tau}$ showers
 are the best suited candidates to produce detectable showers.
Studies of such possibilities were recently presented 
in Refs.~\cite{Bertou:2001vm,Aramo:2004pr,Zas:2005zz,Anchordoqui:2005ey,Miele:2005bt}.

In this paper  to exemplify the potential and the features of the developed simulation tool
we study neutrino induced extensive air showers (EAS) focusing on the fluorescence detector (FD)
set up at the  Pierre Auger Observatory. Taking into account the details of 
the neutrino propagation  inside the Earth and in the Atmosphere as well 
as the topography of the Pierre Auger  Observatory site,
 the directional dependence of the neutrino rate for up-going and down-going neutrino showers 
 (probability maps) are evaluated accounting for the 10\% duty cycle for the FD.  
The WB bound is used as an initial neutrino flux. 
We use  the ellipsoidal model of the Earth based on the so-called datum
WGS84~\cite{wgs84} model. In addition, we use an elevation map of the mountains
surrounding the Auger site. We find that the shape of the surrounding area
influences significantly the calculated  neutrino  rates. To investigate the
response of the FD  for  up-going showers, we generate the longitudinal profiles 
of shower development using the EAS MC generator AIRES~\cite{aires}. The light
propagation  and the hardware detector trigger are simulated by means of  the Auger
software framework called Offline~\cite{Paul:2005}.  
Finally the aperture, acceptance and neutrino rate are calculated for up-going $\nu_{\tau}$ showers.

\begin{figure}[t]
\label{augermap}
\begin{center}
 \includegraphics[width=7.5cm]{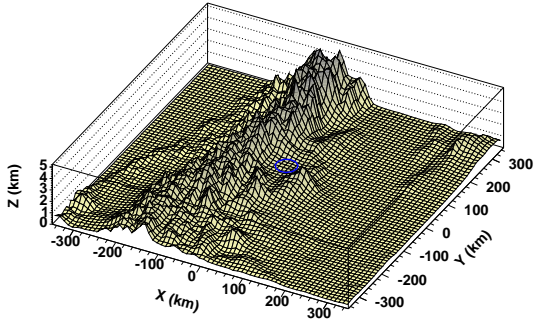}
\end{center}
\vspace*{8pt}
\caption{\label{fig::augermap} Topography of the Auger site according to CGIAR-CSI data. 
The centre of the map corresponds to the centre of the Auger array 
(latitude $\phi_{Auger\_center}=35.25^{\circ}$ S, longitude $\lambda_{Auger\_center}=69.25^{\circ}$ W). The Auger position is marked by a circle. 
}
\end{figure}
\section{Method}
The software tool used in the present paper is based on the code ANIS (All
Neutrino Interaction Simulation)~\cite{Gazizov:2004va}. ANIS is a code
originally developed by A.~Gazizov and M.~P.~Kowalski for the complete
simulation chain of neutrino propagation and interaction in the context of the
AMANDA experiment~\cite{amanda}. It allows to  generate $\nu$-events of all
flavours, to propagate them through the Earth and finally to simulate 
$\nu-$interactions within a specified volume inside the Earth. All relevant
Standard Model processes (charged current (CC), neutral current (NC)  $\nu
N$-interactions and resonant $\bar{\nu}_{e}-e^{-}$ scattering) are
implemented. In addition, neutrino regeneration  at all orders is included. The
density profile of the Earth is chosen according to the Preliminary Earth
Model~\cite{Dziewonski:1981xy}. 
Deep inelastic $\nu - N$-cross-sections are calculated from Quantum Cromo
Dynamics (pQCD) with structure functions according to  CTEQ5~\cite{cteq} and
with a logarithmic extension into the small-x region. The tau decay is
simulated using the program TAUOLA~\cite{Jadach:1993hs}. In order to simulate neutrino showers suitable to be detected by the Pierre
Auger Observatory, some important changes and extensions of the code were
needed. For instance, in ANIS neutrino propagation and tau decay are simulated
only inside the Earth. Instead, we need to simulate also the propagation
through the atmosphere. Moreover, it is necessary to take into account that
the Auger Observatory is positioned on the surface of the Earth. 

First, the topography of the Auger site was implemented. The description of
the relief of the Andes mountains was made according to a digital elevation
map (DEM). These data are available from the Consortium for Spatial
Information (CGIAR-CSI)~\cite{dem}. 
 The implemented map of the area around the Auger site
obtained is shown in Fig.~\ref{fig::augermap}.

\begin{figure}[t]
\label{method}
\begin{center}
 \includegraphics[width=9cm, height=4.5cm]{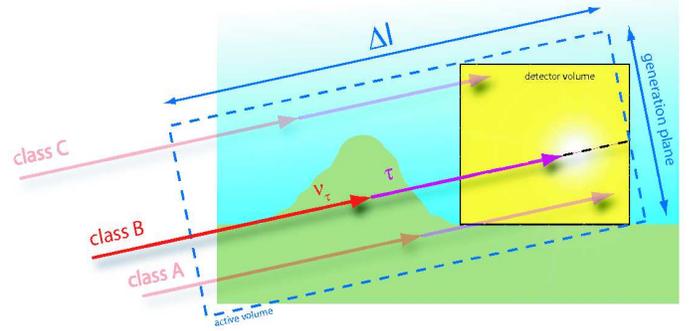}
\end{center}
\vspace*{8pt}
\caption{\label{fig::method}Sketch of the geometry relevant for neutrino simulations, see text for more details. 
}
\end{figure}

Second, a redefinition of the detection volume was done. The original ANIS
version uses the concept of the detection volume~\cite{Kowalski:2000jj}.  
This concept is useful for calculating event rates for a given neutrino flux
$\Phi(E_{\nu},\theta_{\nu})$. The detection volume corresponds to the so
called {\em active volume\/} in which potentially detectable neutrino
interactions are simulated. In the original ANIS this volume is defined as a
cylinder with z-axis parallel to the neutrino direction~\cite{Gazizov:2004va}.
In our adapted version we kept the general idea of {\em active volume\/} 
but with some modifications, as shown in Fig.~\ref{fig::method}.   
As it can be seen from this figure,
the {\em active volume\/} for a given incoming neutrino with energy $E_{\nu}$
is defined by a particular plane $A_{gen}$ and distance $\Delta L$.
The plane $A_{gen}$ is the cross-sectional area of the detector volume and it
was used as reference plane for the generation of incoming neutrinos. 
The area depends on the zenith angle $\theta$ of the incoming neutrino. 
The detector was modelled as a cylinder with radius $R$ and height $H$,
The distance $\Delta L$ is the multiple, $n$, of the average lepton range  
$\left<R_{lep}(E_{lep})\right>$~\cite{Dutta:2000hh}. Inside the {\em active
volume \/}lepton tracks are also  shown. Neutrinos with energy $E_{\nu}$
produce leptons with individual ranges, depending on the fraction of energy
transfered. More precisely, the incoming neutrino is forced to interact in the
{\em active volume \/} according to  its interaction probability  
\begin{equation}
  \label{prob}
  P(E_{\nu},E_{lep},\theta) \simeq N_{A}\times\sigma(E_{\nu}) \times
  \rho(Z)\times \Delta L, 
\end{equation}
where $\sigma(E_{\nu})$ is the total neutrino cross section, $\rho(Z)$  the
local medium density and $N_{A}$ the Avogadro constant.
$P(E_{\nu},E_{lep},\theta)$ is the probability that a neutrino with
energy $E_{\nu}$ crossing the distance $\Delta L$ would produce a lepton with
an energy $E_{lep}$. In this way the production vertex of the lepton is
created. Then, the lepton propagates through the matter, losses some energy
and decays (if it's unstable) in the vicinity of the detector or inside the
detector volume. In order to calculate  physical quantities, one has to weight the events. A
first weight is the interaction probability defined by Eq.~(\ref{prob}). A
second weight comes from the normalisation of the injected neutrino flux 
$\Phi(E_{\nu},\theta_{\nu})$
\begin{equation}
  \label{flux}
  F^{w}_{\nu}=  N_{gen}^{-1} \times \Delta T \times \int_{E_{min}}^{E_{max}}
  \Phi(E_{\nu}) dE_{\nu}   \times \int_{\theta_{min}}^{\theta_{max}}
  A_{gen}(\theta) \times d\Omega, 
\end{equation}
where  $d\Omega=2 \pi \sin \theta d \theta$ is the space angle, $\Delta T$
the observation time and $N_{gen}$ is the number of generated events from
surface $A_{gen}$. Here, we assume the isotropic neutrino flux
$\Phi(E_{\nu},\theta_{\nu})=\Phi(E_{\nu})$. 
The expected event rate of the neutrinos in the detector volume can be
calculated from 
\begin{equation}
\label{nrate}
  N_{ratio}=F^{w}_{\nu} \times \sum_{i=1}^{N_{acc}}  P_{i},
\end{equation}
where $N_{acc}$ is the number of events triggering the detector and passing
all quality cuts of the cascade analysis. 
The normalisation factor $F^{w}_{\nu}$  is chosen such that the rate of the
neutrinos results in the total number of events per year. 

Third,  new tau lepton  decay channels were implemented. In the original ANIS
code the tau lepton decay is simulated using the code  of
TAUOLA~\cite{Jadach:1993hs}. The tau lepton decay is simulated by randomly 
choosing one event from a data base of pre-simulated events.

\begin{figure}[h]
  \begin{center}
\includegraphics[width=8.5 cm,height=5cm]{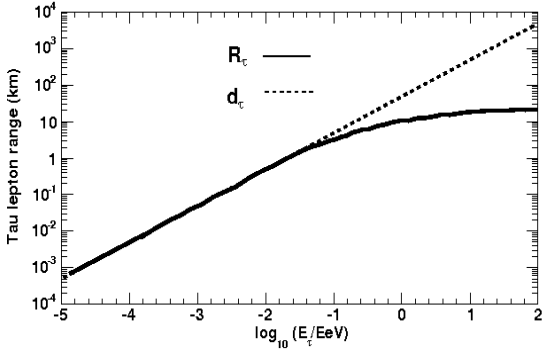}
  \end{center}
 \vspace  {-0.5 cm}
  \caption{\label{fig::taurange} Tau lepton  range  $R_{\tau}$ and its decay
    length $d_{\tau}$ as a function of the energy.}
\end{figure}

 Finally, the tau lepton  decay  routine was  partly rewritten
 to take into account the processes of   tauons escaping  from the Earth's
 surface and decaying  in the atmosphere. 
 In ANIS the tau is propagated in small energy steps until  the age
 of the tau lepton exceeds  the tau lepton lifetime. This procedure works
 quite well if the production and decay vertex are in the same medium (either
 in air or rock), but if the  tau lepton  crosses the Earth's surfaces during
 its  propagation inside the {\em active volume\/},  different amount of
 energy loss in the Earth's crust and air have to  be taken into account. In
 particular, when a tau lepton is generated in the Earth, it  loses energy due
 to ionisation and radiation processes. These energy losses per unit length of
 crossed matter (in g/cm$^2$) 
are usually approximated by  a linear equation (continuous energy loss
 approach), which reads as  
\begin{equation}
\label{dedx}
\frac{dE_{\tau}}{dX}=-\alpha-\beta(E_{\tau}) \times E_{\tau}, 
\end{equation}
where the factor $\alpha$ is due to the ionisation losses and $\beta$ is due to
radiation losses. 
The factor $\alpha$  is negligible for ultra-high energies. 
The factor $\beta$ parameterises the tau lepton energy loss  through
bremsstrahlung, pair production,  
and photonuclear interaction.

Exemplary we have used two  parameterisations:
 $\beta$ is  linear dependent on energy
\begin{equation}
\beta_{A}(E_{\tau})\equiv0.71\times10^{-6} \unit{cm^2 g^{-1}}+
0.35 \times 10^{-18} E  \mbox{ }\unit{cm^{2} g^{-1} GeV^{-1}}
\end{equation}
 used by Aramo et al. in  Ref.~\cite{Aramo:2004pr}, and
\begin{equation}
\beta_{B}(E_{\tau})\equiv(1.508+6.3(E_{\tau}/10^{9})^{0.2})\times 10^{-7}
\unit{cm^{2}g^{-1}}
\end{equation}
from~\cite{Gazizov:2004va,Dutta:2000hh,Tseng:2003pn}.

The lepton range, $R_{\tau}$, is given by the integral of the
inverse of the tau lepton loss rate over the tau lepton energy 
\begin{equation}
\label{range}
R_{\tau}=\frac{1}{\rho(z)}\int \frac{1}{dE/dX} dE. 
\end{equation}
 Thus   differences  in 
$\beta$ will   lead to different  tau lepton ranges (length of the lepton
track) and consequently  influence  the expected neutrino event rate 
and aperture calculations.
 
{ 
 During a single step  the  distance passed by the tau lepton  is evaluated according 
to the following formula
\begin{equation}
\label{range2}
\Delta R_{\tau}=\frac{1}{\rho(z)
  \beta(E^{f}_{\tau})}ln(E_{\tau}^{f}/E_{\tau}^{i}), 
\end{equation}
where  $E^{i}_{\tau}$ is the initial energy of the tau lepton and 
$E^{f}_{\tau}$ is the energy of the  tau lepton after the distance
$\Delta R_{\tau}$. During the energy step 
 $\beta$ is  constant.
}
The distance $\Delta R_{\tau}$ is evaluated with the altitude dependent
density $\rho(z)=\rho^{rock}(z)$ or $\rho(z)=\rho^{air}(z)$.  $\rho^{rock}$ is
the density of the Earth according to the Preliminary Earth
Model~\cite{Dziewonski:1981xy} and $\rho^{air}$ is the air density calculated
according to the  US standard atmosphere (Linsley parameterisation)~\cite{us}.

In Fig.~\ref{fig::taurange} the tau lepton range in standard rock,
$\rho=\unit[2.6]{g/cm^{3}}$ is shown for the parameter set $\beta_{B}$
{ 
based on calculations reported in~\cite{Tseng:2003pn}.
}
One can  see that the  tau lepton range $R_{\tau}$ is of the order of  about
\unit[10]{km} at \unit[1]{EeV}. This value is about 5 times smaller than  the
tau lepton decay length, $d_{\tau}$, at the same  energy. The decay length is
given by $d_{\tau}=c\tau_{\tau} 
(E_{\tau}/m_{\tau})\sim \unit[49]{km} \times (E_{\tau}/\unit[10^{18}]{eV})$ 
for tau lepton, where $c\tau_{\tau}=87.11$ is the tau lepton  lifetime in
$\unit{\mu m}$ 
and  $m_{\tau}=\unit[1777.03]{MeV}$  is  the tau lepton mass~\cite{PDBook}.
Thus, a tau lepton  with an energy of about  $\unit[1]{EeV}$ produced  close to
the Earth's  surface can escape from the Earth before decaying. One can note
also, that  the tau lepton decay length is about \unit[490]{km} for a tau
lepton  energy of about  \unit[10]{EeV}. 
Therefore  an emerging tau lepton can escape  from the atmosphere,
which is assumed to have an height of  \unit[100]{km}.

Applying the neutrino generation code, we obtain the tau lepton decay vertex
position, the energy and momentum  of  the decay products for simulated neutrino showers
with a given energy. 
AIRES was used to generate the longitudinal profiles of charged
particles and the energy deposit based on the ANIS output. A special  mode was
used to inject simultaneously several particles at a given
interaction point. The emission of fluorescence light by the shower together
with its propagation towards the detector and the response of detector itself,
including electronics and trigger, were simulated by the Offline software. 
Finally  on the  basis of  the algorithms implemented in the Offline software
the first two trigger levels called First Level Trigger (FLT) 
and Second Level Trigger (SLT) were simulated~\cite{Abraham:2004dt}. After the
simulated events had passed the FLT threshold trigger (at least one pixel),  a
search for pattern consisting of 5 pixels was performed according to the SLT
algorithm.
In this way  the trigger efficiency of the FD for a given neutrino energy can be defined as
\begin{equation}
\Sigma(E_{\nu})=\frac{N_{SLT}}{N_{Aires}}\times \gamma
\end{equation}
 where $N_{Aires}$ is the the number of AIRES tau showers simulated for a
 given neutrino energy, $N_{SLT}$ the number of showers  passing the  SLT
 condition and $\gamma$ the duty cycle of fluorescence detector.
 
\begin{figure}[h]
  \begin{center}
    \includegraphics[width=8.5 cm,height=6cm, angle=0]{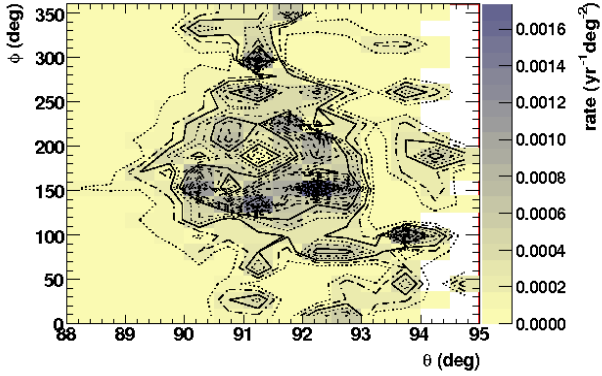}
  \end{center}
\vspace{-0.5 cm}
  \caption{\label{fig::dwratio} Event rate as a function of azimuth and zenith
    for the detector volume in the case of quasi-horizontal $\nu_{\tau}$
    showers, assuming $E_{\tau}^{th}=\unit[1]{ EeV}$, a WB flux, $\Delta L=15
    \left<R_{\tau}(E_{\tau})\right>$, $\beta_{B}$, 10\% duty cycle of FD detector.}
\end{figure}

\section{Results}
\label{Results} 

Simulations  are done for neutrino showers ranging from \unit[1-100]{EeV}.
Usually 500.000 events are generated on the surface $A_{gen}$ 
for different azimuth and zenith angles.  
A cuboid  with an area of  $\unit[50\times60]{km^2}$ and  height of
\unit[10]{km} 
positioned at \unit[1430]{m} a.s.l. was used as detector volume ($\equiv
V_{FD}$). It agrees quite well with the detection volume seen by  FD.

In Fig.~\ref{fig::dwratio} the map of the expected event rate in
$\unit{yr^{-1}}$ as a function of the incoming direction $(\theta,\phi)$ is
shown.  Significant directional differences
in the number of expected events are  seen. The largest rate
is for almost horizontal neutrinos  with the zenith $\theta$ in the range between 
89$^{\circ}$-93$^{\circ}$ and   azimuth $\phi$ in the range between 
100$^{\circ}$  and 270$^{\circ}$ (North-West-South). However,
some peaks in the  rate distribution exist for  azimuth of about
350$^{\circ}$ (East).  
This behaviour is due to the fact that the Auger site is surrounded by the Andes
on the West and smaller but closer mountains in the East (see the  map shown in
Fig.~\ref{fig::augermap}).
\begin{table}[h]
  \caption{\label{tab::updw} Expected event rate in yr$^{-1}$ for
    quasi-horizontal showers. The rate has been evalueated for
    different incoming  $\nu_{\tau}$  directions and at $E^{th}_{\tau}=\unit[1]{EeV}$ 
    assuming the FD detection volume $V_{FD}$.
     The WB neutrino flux, $\beta_{B}$  and 10\% duty cycle of FD detector  have been assumed.}
\begin{center}     
    \begin{tabular}{@{}cccccccc@{}}
      \hline
      \hline
      &  Azimuth      && $N^{FD}_{DW}$  &&        $N^{FD}_{QH}$   \\
      &(deg)          &&  (yr$^{-1}$)   &&         (yr$^{-1}$)   \\
      \hline
      \hline
      Theta    &          && 85$^{\circ}$-90$^{\circ}$  &  &        85$^{\circ}$-95$^{\circ}$&     \\  
      \hline
      East     & -45-45   &&  0.0005 && 0.009  \\
      North    & 45-135   && 0.0007  && 0.012    \\
      West     & 135-225  && 0.0027  && 0.023     \\
      South    & 225-315  && 0.0007  && 0.009     \\
      \hline
      Total    &          && 0.0046  &&0.052     \\
      \hline
      \hline
    \end{tabular}
\end{center}    
\end{table}
On the other words the
correlations between the calculated distribution of the event rate and the
topography of the Auger site is well seen ({\em   mountain
  effect\/}).  The {\em mountain effect \/} is much  more pronounced for $\nu_{\tau}$
  showers very close to the horizon and less obvious for larger zenith angle
  ranges. A quantitative  analysis of this  effect for horizontal  showers is presented  in
  Tab.~\ref{tab::updw} and for up-going showers in Tab.~\ref{tab::UP}.

In Tab.~\ref{tab::updw}, the dependence of the expected number of events from
the direction of incoming neutrinos  is well seen.  For example, in the case
of quasi-horizontal (QH) showers the expected ratio from the West is about 3 times larger than
the ratio  from the  East (for downward-going shower (DW) the  ratio from the
West  is  about 5 times larger than from the East).  Note also that the rate
in the case of QH showers is about one order larger than the rate  for
downward-going  showers.

\begin{table}[h]
\caption{\label{tab::UP} Expected event rate in yr$^{-1}$ for up-going
    neutrino showers for different  ranges of zenith angle for
    $E_{th}=\unit[1]{EeV}$, $\beta_{B}$, FD detector volume $V_{FD}$, WB
    flux and 10\% duty cycle of FD detector. $N^{FD}_{SP}$ is the rate 
    calculated according to the spherical model of the  Earth  with the
    detector positioned  at \unit[10]{m} above the surface.}
   \begin{tabular}{ccccccc}\toprule
      \hline
       & \multicolumn{2}{c}{class A+B+C} & \multicolumn{2}{c}{class A} & Sphere&
       \\ 
      \cmidrule(lr){2-3}\cmidrule(lr){4-5}\cmidrule(lr){6-6}
      $\Theta$ & $N^{FD}$   &  $N_{acc}$ &  $N^{FD}$ & $N_{acc}$
      &$N^{FD}_{SP}$& $\frac{B}{A+B+C}$\\ 
      (deg)         &  (yr$^{-1}$) &            & (yr$^{-1}$) &       &
      (yr$^{-1}$)& (\%)\\ 
      \hline
      \hline
      90-92  & 0.0219$\pm$1\%  & 34205  &0.0108  & 1209  & 0.0105  &48 \\
      92-94  & 0.0168$\pm$2\%  & 9895   &0.0133  & 1248  & 0.0140  &24 \\
      94-96  & 0.0068$\pm$15\%   & 199    &0.0060  & 165   & 0.0059  &11 \\
      96-98  & 0.0008$\pm$43\%   & 29     &0.0008  & 27    & 0.0010  &6  \\
      \hline
      \hline
    \end{tabular}
\end{table}

In Tab.~\ref{tab::UP} the expected event rates for different ranges of zenith
angle are listed. The rate is dominated by events close to the horizon and gets only small
contributions at larger zenith angles. This is due to the fact that 
in the case of up-going  $\nu_\tau$  the regeneration effect plays a important role.
A high-energy $\nu_{\tau}$ interacts in the Earth producing  a tau lepton
which  in turn decay into a $\nu_{\tau}$
with lower energy due to its short lifetime. The regeneration chain
$\nu_{\tau} \rightarrow \tau \rightarrow \nu_{\tau} \rightarrow$ ...
continues until the tau lepton reaches the detector
(in our case the {\em active volume\/}). This effect  leads to a significant
enhancement of the tau lepton  flux  up to  about 40\% more than  the initial
cosmic flux of tau neutrinos of energies   between $10^{6}-10^{8}$ GeV~\cite{Jones:2003zy,Reya:2005vh}.

 Additionally, the rate for the different classes of events defined in
Fig.~\ref{fig::method} is presented. Class A has the production
vertex in the Earth (below horizon) and decay vertices inside $V_{FD}$. The class
B and C have production vertices above the horizon and  decay vertices  inside
$V_{FD}$ (the contribution of class C is about one order smaller than class
B). In principle the difference between these classes are a  measure of the
influence of the {\em mountain effect\/} reflected in a zenith dependent
event rate~\cite{Anchordoqui:2006ey}. As it can be  seen from this Table (the
last column), the contribution of mountains to the total  $\nu_{\tau}$ event
rate is significant. For example,  for nearly horizontal $\nu_{\tau}$ showers
within zenith angle less than 2 degrees below  the horizon, the contribution
is about 50\% while for larger  zenith angle ranges  it is on average less
than 24\%. 
We come to the same conclusion if we estimate the {\em mountain effect\/}
taking into account the calculation  with  the 
 simple  spherical model
of the Earth because the rate calculated in this case agrees very well with the
rate calculated for class~A.

\begin{figure}[h]
 \begin{center}
\includegraphics[width=8.0 cm,height=5cm]{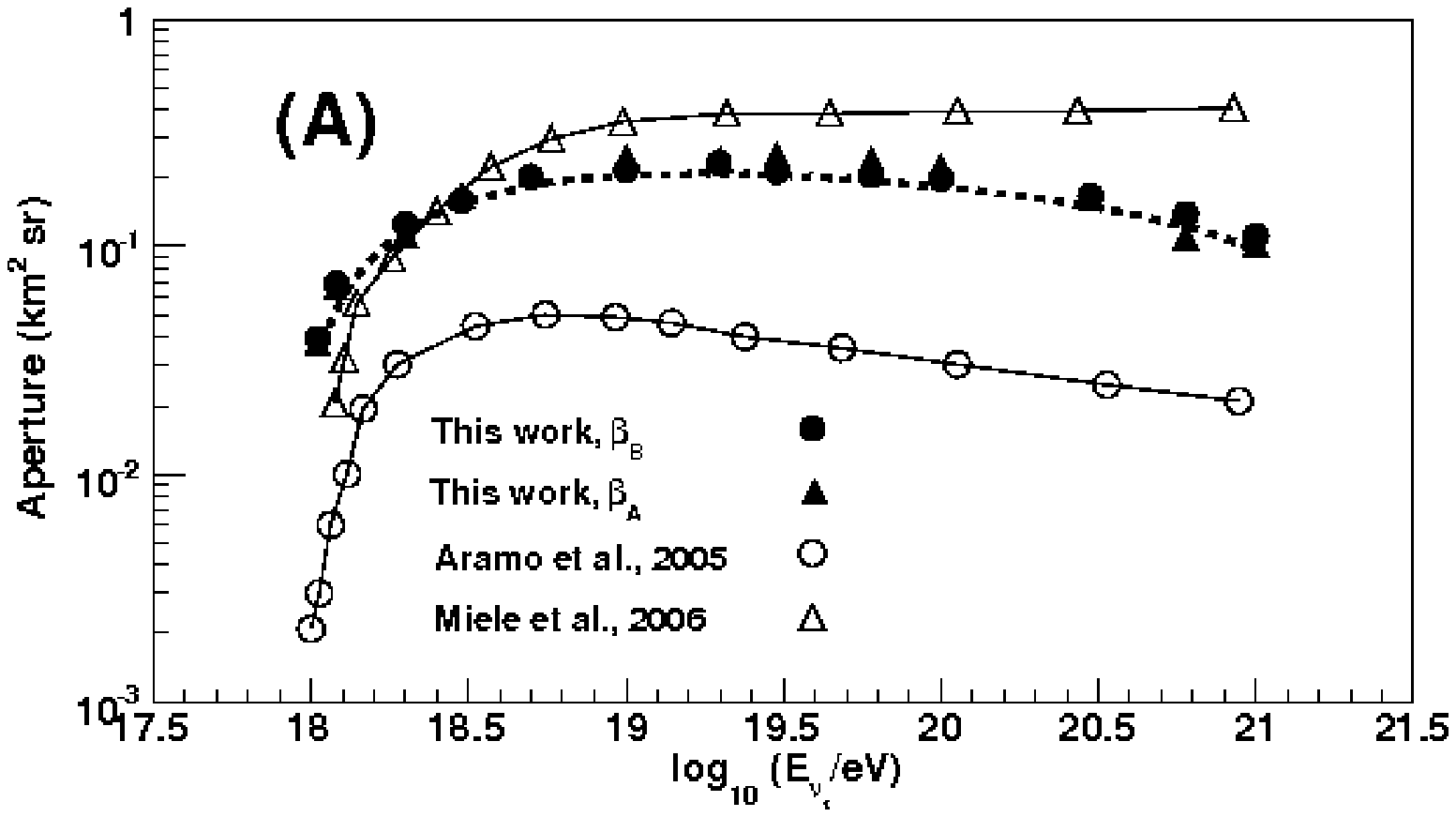}

\includegraphics[width=8.0 cm,height=5.0cm]{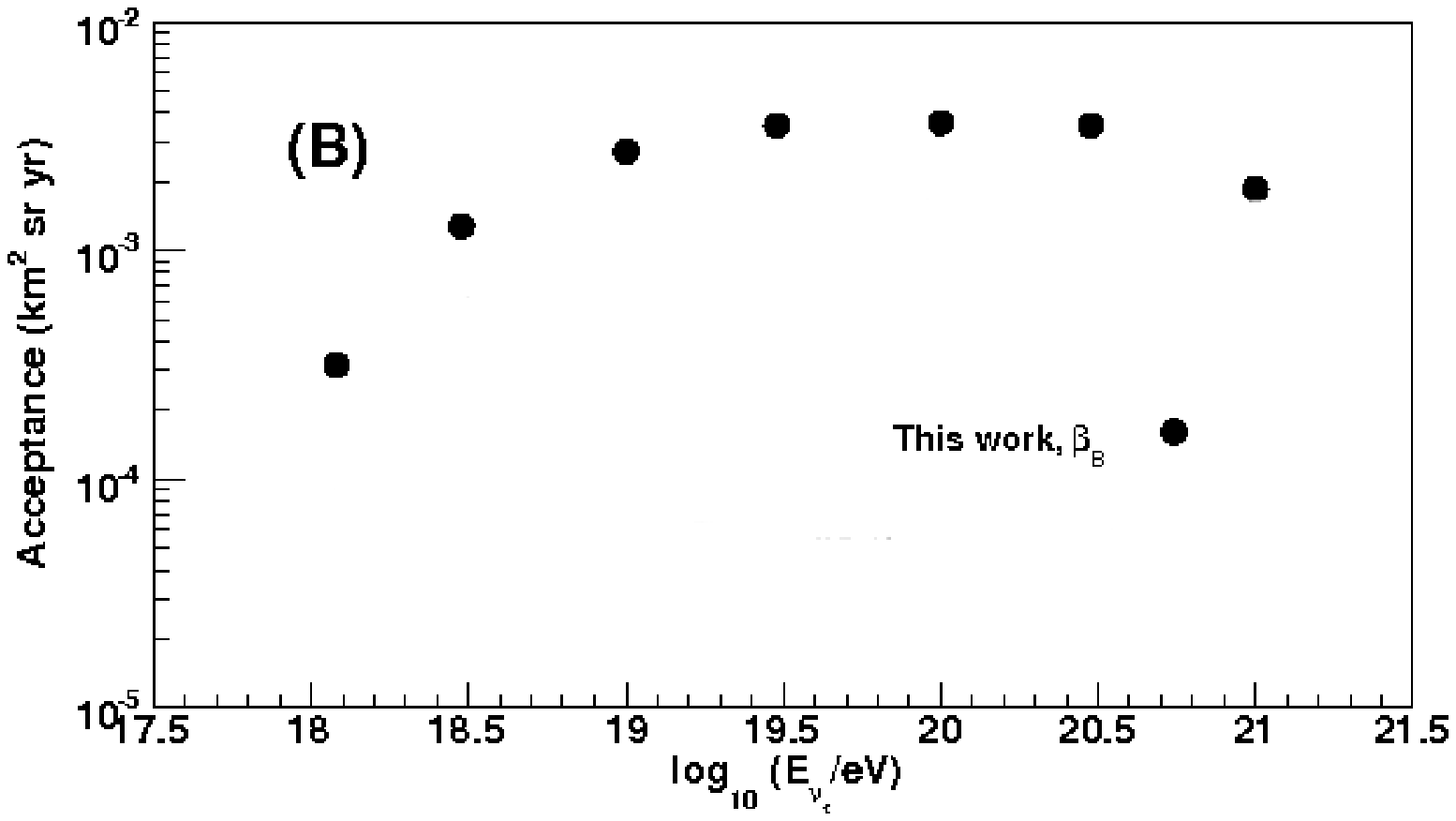}
\end{center}
  \caption{\label{fig::aperture} (A) The  effective   aperture for the Pierre Auger Observatory,
  $A(E_{\nu_{\tau}})$, for $E_{\tau}>10^{18}$  eV, $90^{\circ}<\theta<95^{\circ}$
  ,$0^{\circ}<\phi<360^{\circ}$, $V_{FD}$ and
  $N_{gen}=200000$; (B) The  effective   acceptance $A_{c}(E_{\nu})$ for the Pierre Auger Observatory. The 10\%
  duty cycle of FD  was included.}
\end{figure}

In  Fig.~\ref{fig::aperture}A   the estimation of the aperture for
up-going $\nu_{\tau}$ showers is shown. We define the aperture by
\begin{equation}
A(E_{\nu})=N_{gen}^{-1} \times \sum_{i=1}^{N_{\tau}}
\sum_{j=1}^{N_{\theta,\phi}^{acc}} P_{i,j}(E_{\nu},E_{\tau},\theta) \times
A_j(\theta)\times \Delta \Omega,
\end{equation}
where $N_{\tau}$ is the number of tau leptons with energy $E_{\tau}$.
$E_{\nu}$ is  the initial neutrino energy,  $N_{\theta,\phi}^{acc}$
is the number of tau leptons coming from a given direction inside $V_{FD}$
and  passing our cuts.
The aperture was calculated according to  the  parameterisation of $\beta_{A}$
and  $\beta_{B}$. Different parameterisations lead only to small differences
in the calculated aperture. Additionally, in Fig.~\ref{fig::aperture}A the
aperture calculated by~\cite{Miele:2005bt} is shown. One can see  a fairly
good  agreement within 20\%  between
{
the results presented in this paper
}
and aperture calculated by~\cite{Miele:2005bt} for  energies between
\unit[1]{EeV} and \unit[10]{EeV}.
For larger energies, above \unit[100]{EeV}, our  aperture decreases with
increasing of
neutrino energies. This is due to the fact that for these energies the tau lepton
decay length in air is extremely large (above 5000 km), then the  emerging tau
lepton from the Earth will escape from  the detector, or even from the
atmosphere.

In  Fig.~\ref{fig::aperture}B the FD  acceptance for Pierre Auger Observatory 
is shown. The  acceptance  is calculated from
\begin{equation}
\label{acc}
A_{cc}(E_{\nu})=A(E_{\nu})\times \Sigma(E_{\nu})|_{E_{\tau}> \unit[1]{EeV}}.
\end{equation}
 assuming that a trrigered event is seen at least by one
fluorescence eye of Pierre Auger Observatory. The acceptance  increases from
$\unit[0.001]{km^{2}\, sr\, yr}$ to about $\unit[0.003]{km^{2}\, sr\, yr}$
between \unit[30]{EeV} and  \unit[1000]{EeV}. The acceptance does not saturate
but drops at larger energies since tau leptons at such  energies escape
unseen from the detector volume $V_{FD}$  lowering the acceptance. The
calculated  acceptance differs about  two orders of magnitude from the
calculated  aperture. 
\begin{figure}[t]
  \begin{center}
    \includegraphics[width=8.5 cm,height=5cm, angle=0]{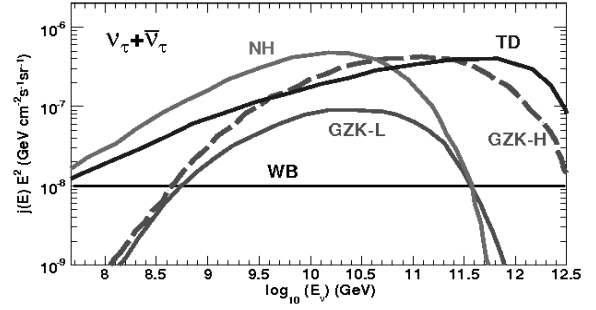}
  \end{center}
  \vspace{-0.5cm}
  \caption{\label{fig::spectra} Tau neutrino and anti-neutrino fluxes
  from various models, taken from Ref.~\cite{Aramo:2004pr}.}
\end{figure}

Finally, in Tab.~\ref{tab::UP2} the rate for different injected neutrino fluxes  shown in Fig.~\ref{fig::spectra},  
based on our acceptance and apperture calculation, are listed.
The WB  rate is obtained for Waxman-Bahcall bound. In this paper
 we use the conservative estimate of this bound
$\Phi(E_{\nu_{\tau}+\bar{\nu}_{\tau}})=1\times
10^{-8} E^{-2} \mbox{ (GeV s$^{-1}$ cm$^{-2}$ sr)}$.  
Other rates are reported for the same neutrino fluxes considered in section 2 of 
Ref.~\cite{Aramo:2004pr}. The two GZK fluxes refer to the two possible scenarious 
for cosmogenic neutrinos, which are those produced from an initial flux of UHE protons.
Instead the NH (New Hadrons) and TD (Topological Defects) cases  are two examples
of exotic models. As one can see from Tab.~\ref{tab::UP2}, the number of expected
events equals about 1 neutrino event  per 2 years in the case of  our aperture calculation. Taking
into account the FD detector properties rate is  about one order smaller, 
but neutrino event  still can be observable by Auger detector during a few decade of operational. 

\begin{table}
\caption{\label{tab::UP2} Expected event rate in yr$^{-1}$ for the different neutrino fluxes
based on our aperture ($N_{Aper}$)  and 
acceptance  calcultion ($N_{Acc}$). Results for   $\nu_{\tau}$ events with  zenith angle in the range $90^{\circ}-95^{\circ}$, 
assuming $E_{\tau}^{th}=\unit[1]{ EeV}$, a WB flux, $\Delta L=15
\left<R_{\tau}(E_{\tau})\right>$, $\beta_{B}$, 10\% duty cycle of FD detector.}
{\begin{tabular}{@{}cccccccccccc@{}} \toprule
      \hline
      &          &      & WB &    &   GZK-L&GZK-H &&  TD &  NH\\
      \hline
      & $N_{Aper}$ &   &0.044 &&0.456&0.699&& 0.546& 1.478\\
      & $N_{Acc}$ &  & 0.004 && 0.042& 0.072 && 0.053& 0.137\\

      \hline
      \hline
    \end{tabular}}
\end{table}
Note that event rate becomes larger for lower threshold energy. In fact FD
detector can record also the shower with enegies above $10^{17}$ eV.

\section{Summary}
\label{Summary} 

A tool to simulate neutrino propagation and interaction in the Earth and in the
atmosphere taking into account the local topographic conditions, was presented. 
A detailed investigation of neutrino 
induced showers was performed. 
The focus was put on the neutrino  sensitivity  for the
FD detector of the Pierre  Auger Observatory and therefore we have used the 
digital elevation map for this site. The probability map (event rate
distribution) and the  aperture were calculated and the impact of the mountains
surrounding the Auger site was quantified. General features are consistent
with previous results, 
but as a consequence of the surrounding mountains
a pronounced variation of the event rate as a function of the azimuth is seen.

\section*{Acknowledgment}
We acknowledge the financial support by the HHNG-128 grant of the
Helmholtz foundation. 

\IEEEtriggeratref{11}

\end{document}